\documentstyle[12pt]{article}
\def\pmb#1{\setbox0=\hbox{#1}%
\kern.0em\copy0\kern-\wd0
\kern-.04em\copy0\kern-\wd0
\kern.08em\copy0\kern-\wd0
\kern-.04em\raise.0433em\box0 }         

\newcommand{\nc}{\newcommand}
\nc{\pek}[1]{\cite{#1}}
\nc{\enr}[1]{(\ref{#1})}
\nc{\kal}[1]{{\cal{#1}}}

\def\bop#1{\setbox0=\hbox{$#1M$}\mkern1.5mu
        \vbox{\hrule height0pt depth.04\ht0
        \hbox{\vrule width.04\ht0 height.9\ht0 \kern.9\ht0
        \vrule width.04\ht0}\hrule height.04\ht0}\mkern1.5mu}

\begin{document}

\newcommand{\inv}[1]{{#1}^{-1}} 

\renewcommand{\theequation}{\thesection.\arabic{equation}}
\newcommand{\beq}{\begin{equation}}
\newcommand{\eeq}[1]{\label{#1}\end{equation}}
\newcommand{\ber}{\begin{eqnarray}}
\newcommand{\eer}[1]{\label{#1}\end{eqnarray}}
\begin{center}

                                \hfill    hep-th/0101213\\
                                \hfill    USITP-01-03\\
                                \hfill    IHP-2000/29\\
                                \hfill    Jan. 2001

\vskip .3in \noindent

\vskip .1in

{\large \bf {Strings at the Tachyonic Vacuum}}
\vskip .2in

{\bf Ulf Lindstr\"om}$^a$\footnote{e-mail address: ul@physto.se}
 and  {\bf Maxim Zabzine}$^{ab}$\footnote{e-mail address: zabzin@physto.se} \\


\vskip .15in

\vskip .15in
$^a${\em  Institute of Theoretical Physics,
University of Stockholm \\
Box 6730,
S-113 85 Stockholm SWEDEN}\\
\vskip .15in
$^b${\em Institute Henri Poincar\'e 11, rue P. et M. Curie \\
 75231 Paris Cedex 05, France}

\bigskip


 \vskip .1in
\end{center}
\vskip .4in
\begin{center} {\bf ABSTRACT } 
\end{center}
\begin{quotation}\noindent 
  We study the world-volume effective action of Dp-brane at the tachyonic
 vacuum which is equivalent to the zero tension limit. 
 Using the Hamiltonian formalism we discuss the algebra of constraints and show
 that there is a non-trivial ideal of the algebra which corresponds to Virasoro like
 constraints.
 The Lagrangian treatment of the model is also considered. 
 For the gauge fixed theory we construct the important subset of 
 classical solutions which is equivalent to the string theory solutions in  
 conformal gauge.
 We speculate on a possible quantization of the system.
   At the end a brief discussion  of 
 different background fields  and fluctuations around the tachyonic vacuum
 is presented.    
\end{quotation}
\vfill
\eject

\section{Introduction and motivation} 

 The problem of tachyon condensation is an old subject in the context of 
 the string theory \cite{Halpern}. 
 Recently it has been conjectured that the tachyonic vacuum in open string 
theory on the D-brane
 describes the closed string vacuum without D-branes and that various soliton
 solutions of the theory describe D-branes of lower dimension \cite{Sen:1999mh}.      
  This conjecture has been supported by number calculations within the first and
 second quantized string theory.

 One can get some insight into the problem by considering the world-volume effective
 action \cite{Sen:1999md} which describes the D-brane around the tachyonic vacuum.  
 Recently there has been some effort directed towards
  identifying  of string-like classical  solutions whose tension 
 matches that of fundamental string \cite{yi, Bergman:2000xf, Harvey:2000jt,
 Gibbons:2000hf, Sen:2000kd}. 
 In fact the D-brane action at the tachyonic 
 vacuum is equivalent to the zero tension limit of the D-brane action. 
 The zero tension limit
 was with a different motivation studied previously 
 in \cite{Lindstrom:1997uj, Gustafsson:1998ej, Lindstrom:1999tk} 
 and a string picture was obtained there as well.

 In this short note we would like to revise and clarify certain arguments 
 from  \cite{Lindstrom:1997uj, Gustafsson:1998ej, Lindstrom:1999tk} 
 in light of the
 new motivation. Unlike \cite{Gibbons:2000hf, Sen:2000kd}
 we do not start from the gauge fixed model.
 The ultimate goal is to relate two different theories (zero tension
 D-brane theory and string theory) to each other. These two theories have 
 different gauge symmetries and are unrelated a priori.
  We study the algebra of
 constraints of D-brane theory at the tachyonic vacuum in detail. 
 It turns out that fundamental string-like Hamiltonian constraints generate
 a subalgebra (in fact an ideal) of the full algebra
 of the model at the tachyonic vacuum and that 
 there is natural embedding of this subalgebra provided by the ``electric flux''.
 We hope that the present analysis will clarify the general situation
 and as well as explore the relation between the static gauge results 
 \cite{Gibbons:2000hf,
  Sen:2000kd} and the general situation. Our formalism is 
 Poincar\'e covariant, i.e., we avoid the gauge fixing which breaks
 the Poincar\'e invariance (i.e, static gauge).     

 In addition we study the Lagrangian which describes the dynamics of D-brane
 at the tachyonic vacuum. In the set of solutions of the classical equations of 
  the gauge fixed model we identify the subset of solutions that correspond 
 to string theory
 solutions in conformal gauge. We argue that the model might be consistently 
 quantized  if we regard the electric and magnetic fields as a type  of
 background fields.   

 At the tachyonic vacuum the RR background fields decouple completely from
 the dynamics since the Wess-Zumino couplings are proportional to the derivative
 of the tachyon field. This is natural
 since one expects that the tachyonic vacuum is equivalent to the closed
 string vacuum without D-branes. However the fluctuations around the vacuum should
 describe the D-branes of lower dimension. Therefore it is natural 
 to study these fluctuation within the present framework. 
  At the end of the paper we thus briefly discuss small fluctuation around the
  tachyonic vacuum. 

 The paper is organized as follows: in Section 2 we study the algebra of constraints 
 within the Hamiltonian formalism. Section 3 is devoted to the Lagrangian treatment
 of the model and to the classical solutions of the gauge fixed theory.
 In the Section 4 the role of the background antisymmetric tensor fields
 and small fluctuations around the vacuum are discussed. In the last Section 
 we summarize the results and propose some future research.     

\section{Hamiltonian treatment of BI theory}

 In this section we study the algebra of constraints in detail and
 find that the string-like constraints generate an ideal
 inside the full algebra.

 Let us start by considering the effective D-brane action \cite{Sen:1999md} 
 with constant tachyon field $T$
\beq
S = T_p V(T) \int d^{p+1} x \,\,
 \sqrt{-det(\gamma_{\alpha\beta} + 2\pi\alpha'F_{\alpha\beta})}
\eeq{b1}
 where $\gamma_{\alpha\beta} = \partial_\alpha X^\mu \partial_\beta X^\nu G_{\mu\nu}$
 is the pullback of the space-time metric, $F_{\alpha\beta} = \partial_{[\alpha}
 A_{\beta]}$ is the field strength for the $U(1)$ gauge field $A_\alpha$ and 
 $V(T)$ is a tachyon potential. For the moment 
 we ignore the antisymmetric background tensor fields (NS two-form and RR forms).
 In what follows we assume that the D-brane is closed (i.e., that the
  appropriate periodicity
 conditions  on the fields are imposed) or that there is appropriate fall off of
 the field at the spatial infinity. This assumption is needed to avoid
  possible boundary terms. 

 Since (\ref{b1}) is a generally covariant system the naive Hamiltonian vanishes. 
 The constraints can be straightforwardly derived from the action (\ref{b1}) 
 \cite{Lindstrom:1997uj}
\begin{eqnarray}
{\cal H}_a& = & P_{\mu} \partial_a X^\mu + \pi^b F_{ab},\label{b2}\\
{\cal H} &=&  P_\mu G^{\mu\nu}  P_\nu + \frac{1}{(2\pi\alpha')^2}
 \pi^a \gamma_{ab} \pi^b 
 + T_p^2 (V(T))^2 det (\gamma_{ab} +2\pi\alpha' F_{ab}),\label{b3}\\
{\cal G}&=& \partial_a \pi^a,\,\,\,\,\,\,\,\,\,\,\,\,\,\,\pi_0=0\label{b4}. 
\end{eqnarray}
 where we use lower case Latin letters for the  spatial indices.
 There are $(p+3)$ constraints as there should be corresponding to
  the $(p+1)$ diffeomorphisms and $U(1)$ symmetries. 
 It is convenient to  smear the constraints with test functions
$$ {\cal H}_a[N^a] = \int d^p x\,N^a(x) {\cal H}_a(x),\,\,\,\,\,\,\,\,\,
 {\cal H} [M] = \int d^p x M(x) {\cal H}(x)$$
\beq
 {\cal G}[\Lambda] = \int d^p x \Lambda(x) {\cal G}(x)
\eeq{b4a}
 where $N^a$ is a p-dimensional vector, $\Lambda$ is scalar and
 $M$ is scalar density of weight minus one. $M$ has this
 weight because the constraint (\ref{b3}) transforms as a scalar density
 of weight  two and we would like to write the Hamiltonian
 as a sum of the constraints smeared by test functions.
In analogy with general relativity we call $N^a$ and $M$ shift vector and lapse 
 function, respectively. As usual for gauge theories we can identify $\Lambda$
 with the zero component of the gauge vector potential. 
 The constraints obey an algebra whose non-zero brackets are
\begin{eqnarray}
 \{ {\cal H}_a[N^a], {\cal H}_b[M^b] \}& =& {\cal H}_a[ {\cal L}_{\vec{N}} M^a]
   +  {\cal G} [N^a M^b F_{ba}],\label{FA1}\\
 \{ {\cal H}_a [N^a], {\cal H}[M] \}& =& {\cal H} [{\cal L}_{\vec{N}} M ]
 + {\cal G} [2M \pi^a \gamma_{ab} N^b],\label{FA2}\\
 \{ {\cal H}[N], {\cal H}[M] \}& =& {\cal H}_a [4(\pi^a\pi^b+A^{(ab)})(N\partial_bM-
 M\partial_b N)] \label{FA3},
\end{eqnarray}
 with the following notation 
\beq
 A^{ab} = T_p^2 V(T)^2 \frac{1}{(p-1)!}\epsilon^{aa_1...a_{p-1}} 
\epsilon^{bb_1 ...b_{p-1}}
 (\gamma_{a_1b_1} + F_{a_1b_1})...(\gamma_{a_{p-1}b_{p-1}} +  F_{a_{p-1}b_{p-1}}) 
\eeq{Aab}
 and $A^{(ab)}=\frac{1}{2}(A^{ab}+A^{ba})$. For the time 
 being we drop a factor $(2\pi\alpha')$
 to avoid cluttering the formulae.  This factor can be easily 
 restored in the final expressions.
 The full Hamiltonian is given by
\beq
H= {\cal H}_a[N^a] + {\cal H}[M] + {\cal G}[\Lambda],
\eeq{hamil}
 and provides the time evolution of the fields. The algebra (\ref{FA1})-(\ref{FA3})
 is not closed and has field dependent structure constants.
 As far as the authors are aware the classical BRST charge for this system 
 has not been constructed. By analogy to the p-brane 
 case \cite{Henneaux:1983um} one may expect that the BRST 
 charge will have quite a high rank (maybe $p$
 as for $p$-branes \cite{Henneaux:1983um}). Therefore it seems
 difficult to quantize this model from  first 
 principles\footnote{Even though we call the BI theory the effective theory it is 
 still an unsettled
 question what is more fundamental, strings or D-branes. Thus 
  a possible consistent quantization of the BI theory is an important issue.}.  

 If the tachyon field is frozen at its minimum $T=T_0$ ($V(T_0)=0$) the 
 constraints are reduced to the following set 
\beq
{\cal H}_a  =  P_{\mu} \partial_a X^\mu + \pi^b F_{ab},
\eeq{D3mom}
 and
\beq
{\cal H} = P_\mu G^{\mu\nu} P_\nu + \frac{1}{(2\pi\alpha')^2} 
 \pi^a \gamma_{ab} \pi^b, 
\eeq{D3ham}
 together with Gauss' law $\partial_a \pi^a =0$ and $\pi_0=0$. The constraints 
 (\ref{D3mom})-(\ref{D3ham}) obey the following algebra
\begin{eqnarray}
 \{ {\cal H}_a[N^a], {\cal H}_b[M^b] \}& =& {\cal H}_a[ {\cal L}_{\vec{N}} M^a]
   +  {\cal G} [N^a M^b F_{ba}],\label{ATV1}\\
 \{ {\cal H}_a [N^a], {\cal H}[M] \}& =& {\cal H} [{\cal L}_{\vec{N}} M ]
 + {\cal G} [2M \pi^a \gamma_{ab} N^b], \label{ATV2}\\
 \{ {\cal H}[N], {\cal H}[M] \}& =& {\cal H}_a [4\pi^a\pi^b(N\partial_bM-
 M\partial_b N)]. \label{ATV3}  
\end{eqnarray}
 The algebra (\ref{ATV1})-(\ref{ATV3}) is similar to that of the 
 full theory (\ref{FA1})-(\ref{FA3}). In fact  at the tachyonic
 vacuum the system  has
 the gauge symmetries of the full theory\footnote{This is not the case in the
 $\alpha' \rightarrow \infty$ limit, where the gauge algebra has a
 different form.}.
  For the usual p-branes (i.e., with
 the gauge field in (\ref{b1}) equal to zero) this is not the case. The zero tension
  algebra for a p-brane has a completely  different form from that of
  the full algebra. 
  
 There is one important difference in the field dependent
 structure constants of (\ref{ATV3}) and (\ref{FA3}) which plays a key role in finding 
 the string like subalgebra (\ref{subaTV1})-(\ref{subaTV3}).
 Explicitly the right hand side of (\ref{ATV3}) is
\beq
{\cal H}_a [4\pi^a\pi^b(N\partial_bM- M\partial_b N)] = \int d^px\,
4\pi^a\pi^b(N\partial_bM 
 - M\partial_b N) ( P_{\mu} \partial_a X^\mu + \pi^c F_{ac}), 
\eeq{LHATV3}
 where the last term vanishes identically because of the antisymmetry of $F_{ab}$.
 Thus the right hand side of (\ref{LHATV3}) leads to
 $P_\mu \pi^a \partial_a X^\mu=0$ showing that $\pi^a \partial_a X^\mu=0$ is a
 ``preferred'' direction on the world-volume. 
   
 As a direct result of (\ref{D3mom})-(\ref{D3ham}) there is not
 much dynamics for the $U(1)$ degrees of freedom. 
 For instance, the momenta $\pi^a$ satisfy the following equations
\beq
 \partial_a \pi^a =0,\,\,\,\,\,\,\,\,\,\dot{\pi}^a = {\cal L}_{\vec{N}} \pi^a,
\eeq{dynpi}
 and thus $\pi^a$ completely decouples from the other fields, except from
 the Lagrangian multiplier, $\vec{N}$, (the shift vector).    
 Because of (\ref{LHATV3}) one may decompose the constraints
 (\ref{D3mom})-(\ref{D3ham})
 in the following fashion
\beq
{\cal H}_\pi = P_{\mu}\pi^a \partial_a X^\mu,\,\,\,\,\,\,\,\,\,\,\,\,\,\,\, 
{\cal H} = P_\mu G^{\mu\nu} P_\nu + \frac{1}{(2\pi\alpha')^2} 
 \pi^a \gamma_{ab} \pi^b, 
\eeq{deccTV}
 the other $(p-1)$ generators being 
\beq
 {\cal H}^{\perp}_a [\tilde{N}^a] = \int d^p x\,\tilde{N}^a (P_\mu \partial_a X^\mu
 + \pi^b F_{ab}) ,
\eeq{leftgen}
 which generate the general coordinate transformations with parameter
 $\tilde{N}^a$ (see appendix).
 The decomposition (\ref{deccTV})-(\ref{leftgen}) 
 may be thought of as a decomposition of the shift vector $N^a$ along 
 directions ``parallel'' and ``orthogonal'' to  $\pi^a$ 
\beq
 N^a = N \pi^a + \tilde{N}^a ,
\eeq{decomplapse}
 where $N$ transforms as density of weight minus one. 
 To really implement the concept ``parallel'' and ``orthogonal'' involves
 a metric, e.g., (the spatial part of) the induced metric, and leads to
 unwanted additional constraints. All we need is for $N\pi^a$ and $\tilde{N}^a$
 to be linearly independent. 
 The details of this decomposition
 are not essential in what follows, however. The appropriate decomposition of the 
  constraint  ${\cal H}_a$ can  always be done locally.
  At the global level  
 (\ref{decomplapse}) may imply  restrictions on the topology of the D-brane
 world-volume. In the following discussion we will disregard this potential 
 complication. 
  
 For the given $\pi^a$ we decompose the full set constrains as in 
 (\ref{deccTV})-(\ref{leftgen}).
 The algebra of (\ref{deccTV}) is given by  
\begin{eqnarray}
 \{ {\cal H}_\pi[N], {\cal H}_\pi[M] \}& =& {\cal H}_\pi[ \pi^a (N\partial_a M 
 -M\partial_a N)],\label{subaTV1}\\
 \{ {\cal H}_\pi [N], {\cal H}[M] \}& =& {\cal H} [4\pi^a(N\partial_a M -M\partial_a N)
 + NM(\partial_a \pi^a)], \label{subaTV2}\\
 \{ {\cal H}[N], {\cal H}[M] \}& =& {\cal H}_\pi [4\pi^a (N\partial_a M-
 M\partial_a N)] . \label{subaTV3}  
\end{eqnarray}
 This subalgebra closely resembles the algebra of constraints of the Nambu-Goto
 string. We have thus found a non trivial embedding of one algebra into another with
  field dependent structure constants.

 Guided by this we introduce the following constraints
\beq
Q^{\pm}[N] = {\cal H}[N]\pm2{\cal H}_{\pi}[N] = \int d^px\,N (P_\mu\pm 
 G_{\mu\sigma}\pi^a\partial_a X^\sigma)G^{\mu\nu}(P_\nu\pm 
 G_{\nu\rho}\pi^b\partial_b X^\sigma) ,
\eeq{notaQ}
 which are analogs of Virasoro constraints.
 In terms of the new constraints the algebra (\ref{ATV1})-(\ref{ATV3}) becomes
\begin{eqnarray}
 \{  Q^{+}[N], Q^{+}[M] \}& =& Q^{+}[ 8\pi^a (N\partial_a M 
 -M\partial_a N)],\label{QQplus}\\
 \{ Q^{-} [N], Q^{-}[M] \}& =& Q^{-} [8\pi^a(N\partial_a M 
 -M\partial_a N)], \label{QQminus}\\
 \{ Q^{+}[N], Q^{-}[M] \}& =& Q^{+}[NM\partial_a\pi^a] +
 Q^{-}[NM\partial_a\pi^a] , \label{QQmixed} \\
 \{ {\cal H}^{\perp}_a [\tilde{N}^a], Q^{\pm}[M] \}& =& 
  Q^{\pm} [{\cal L}_{\vec{\tilde{N}}} M] +
  {\cal G}[2\tilde{N}^a\pi^b M\gamma_{ab}] , \label{QQminus1}\\
 \{ {\cal H}^{\perp}_a [\tilde{N}^a], {\cal H}^{\perp}_b [\tilde{M}^b] \}& =& 
 {\cal H}^{\perp}_a 
 [{\cal L}_{\vec{\tilde{N}}} \tilde{M}^a]+ {\cal G}[\tilde{N}^a \tilde{M}^b
 F_{ba}] . \label{QQminus2}
\end{eqnarray}
 This algebra is thus exactly the same as (\ref{ATV1})-(\ref{ATV3}) but written 
 in a different form. The relation (\ref{QQmixed}) can also be written as
\beq
\{ Q^{+}[N], Q^{-}[M] \} = {\cal G}[2NM(P_\mu G^{\mu\nu}P_\nu +
 \pi^a \gamma_{ab}\pi^b)]. 
\eeq{difQQidff}
 The algebra (\ref{QQplus})-(\ref{QQminus2}) follows straightforwardly
  from the previous calculations.
  The only relation which needs checking is 
\beq
 \{ {\cal H}_a[N^a], {\cal H}_\pi [M] \} = {\cal H}_\pi [{\cal L}_{\vec{N}}M], 
\eeq{HHpi}
 where there are no restrictions on $N^a$. 
 The algebra (\ref{QQplus})-(\ref{QQminus2}) contains  the Virasoro 
 like generators $Q^{\pm}$  which together with Gauss law ${\cal G}$ 
 generate an ideal of the full algebra. We know of no other  
 nontrivial ideal of a gravity algebra with field dependent structure
 constants. It is unclear how this ideal
 is manifested in  the BRST charge and other gauge theory quantities, but the existence 
 of a nontrivial ideal may perhaps throw some light on the relation
 between  theories with different gauge symmetries. We hope to return to
 this question elsewhere.

 The algebra (\ref{QQplus})-(\ref{QQminus2}) is not closed and
  has field dependent structure constants. However since $\pi^a$
 decouples (see (\ref{dynpi})) it is tempting to 
  assume that Gauss' law holds strongly;
 $\partial_a \pi^a = 0$. 
 Thus one may regard it as an 
 ``background field''. Thus, considering  a definite field configuration $\pi^a$ with   
  $\partial_a \pi^a = 0$  we may study the behavior of the system
 with this given ``electric flux'' $\pi^a$.  
 Introducing a mode expansion for $\pi^a$ and a mode expansion of the constraints
 $Q^{\pm}$, ${\cal H}^{\perp}_a$ 
\beq
 \pi^a=\sum\limits_{\vec{N}} \pi^a_{\vec{N}} e^{-i\vec{N}\vec{x}},\,\,\,\,\,\,\,\,\,\,
 L^{\pm}_{\vec{M}} =Q^{\pm}[\frac{1}{8}e^{-i\vec{M}\vec{x}}],\,\,\,\,\,\,\,\,\,\,
 H_{a,\vec{N}}={\cal H}^{\perp}_a[e^{-i\vec{N}\vec{x}}]
\eeq{expQHpi} 
 we get the following classical algebra
\begin{eqnarray}
 \{  L^{+}_{\vec{N}}, L^{+}_{\vec{M}} \}& =& i \sum\limits_{\vec{S}} 
\pi^a_{\vec{S}} (N_a -M_a)
 L^{+}_{\vec{N}+\vec{M}+\vec{S}},\label{LLplus}\\
\{  L^{-}_{\vec{N}}, L^{-}_{\vec{M}} \}& =& i \sum\limits_{\vec{S}} 
\pi^a_{\vec{S}} (N_a -M_a)
 L^{-}_{\vec{N}+\vec{M}+\vec{S}},\label{LLminus}\\
\{  L^{+}_{\vec{N}}, L^{-}_{\vec{M}} \}& =& 0,\label{LLplusminus}\\
 \{ H_{a,\vec{N}}, L^{\pm}_{\vec{M}} \} &=&i(N_a-M_a)
 L^{\pm}_{\vec{N}+\vec{M}},\label{HL}\\ 
\{ H_{a,\vec{N}}, H_{b,\vec{M}} \} &=& i N_b H_{a,\vec{N}+\vec{M}} -
 iM_a H_{b,\vec{N}+\vec{M}}.\label{HHH}
\end{eqnarray}
 We see that the subalgebra generated by $L^{+}_{\vec{N}}$ and $L^{-}_{\vec{N}}$
 is an ideal of the whole gauge algebra.

 The subalgebra (\ref{LLplus}) (as well as (\ref{LLminus})) can be thought of 
 as generalizations of the Virasoro algebra. To illustrate this let us
 choose $\pi^a$ to be constant and thus the subalgebra (\ref{LLplus})
  to be
\beq
\{  L^{+}_{\vec{N}}, L^{+}_{\vec{M}} \} =  i \pi^a (N_a -M_a)
 L^{+}_{\vec{N}+\vec{M}}. 
\eeq{LLplusnew}
 Generically this algebra contains  
 $p$ copies of the standard Virasoro algebra
$$\{  L^{+}_{(n,0,0,...,0)}, L^{+}_{(m,0,0,...,0)} \} =  i \pi^1 (n - m)
 L^{+}_{(n+m,0,0,...,0)},$$
\beq
\{  L^{+}_{(0,n,0,...,0)}, L^{+}_{(0,m,0,...,0)} \} =  i \pi^2 (n - m)
 L^{+}_{(0,n+m,0,...,0)},
\eeq{VirasLLplus2}
$$....$$
  $\pi$ can be absorbed into a redefinition of the generators
 to bring these to the standard Virasoro algebra form.
 The embedding of the Virasoro algebras depends on
 the relative orientation of $\pi^a$ in $R^p$. Thus at the level of 
 the classical gauge algebra we see that  there is a string sector of the 
 D-brane at the tachyonic vacuum.

In the quantum theory the algebra 
 (\ref{LLplus})-(\ref{HHH}) would have a central extension which
 should be related to the ``electric flux'' $\pi^a$. Thus a consistent quantization 
 of the system may impose restrictions on the allowed ``electric fluxes''.
  Since (\ref{LLplus})-(\ref{HHH}) is a standard Lie algebra the classical BRST 
 charge can be constructed and it will have rank one. Therefore in 
 principle one may quantize  the system. 
 However, the relation of the BRST charge for (\ref{LLplus})-(\ref{HHH})
  to the full BRST charge of the system (\ref{QQplus})-(\ref{QQminus2}) remains to
 be determined.  
In the next section we propose another way of quantizing the system 
 where  an algebra similar to  (\ref{LLplus})-(\ref{HHH})
  appears as the algebra of residual symmetries of the model, after a partial
 gauge-fixing.

\section{Lagrangian analysis}

 In this section we would like to review the problem from  the Lagrangian 
 point of view. We learn that all, as is usually the case, results 
 can be obtained from the  Lagrangian approach 
 without any direct reference to Hamiltonian analysis.
 
 The Lagrangian (\ref{b1}) is not suited for freezing the tachyon field to
 its minimum (or taking the zero tension limit). Thus the natural approach is to rewrite 
 the Lagrangian (\ref{b1})  in a different but classically equivalent form which is 
  appropriate for the limit in.  
 Following calculation in \cite{Lindstrom:1997uj} one constructs the following
 equivalent action  for the model
$$ S=  \int d^{p}x\, [(E_1^\alpha E_1^\beta - E_2^\alpha E_2^\beta)
 \gamma_{\alpha\beta} + 2\pi\alpha' E_{[2}^\alpha E_{1]}^\beta F_{\alpha\beta} -$$
\beq
-\frac{1}{(2E_1^0)^2}
 T_p^2 (V(T))^2 \det (\gamma_{ab}+2\pi\alpha' F_{ab})].
\eeq{exaction}
 Eliminating  $E_1^\alpha$ and $E_2^\alpha$ gives back the
 ``classical'' BI action (\ref{b1}). 
 To study a D-brane at the tachyonic vacuum  we drop the last term 
 in (\ref{exaction})
\beq
 S= \int d^{p}x\, [(E_1^\alpha E_1^\beta - E_2^\alpha E_2^\beta)
 \gamma_{\alpha\beta} + 4\pi\alpha' \partial_\alpha(E_{[1}^\alpha E_{2]}^\beta)A_\beta]. 
\eeq{actionA}
 This action corresponds to the Hamiltonian constraints 
 (\ref{D3mom})-(\ref{D3ham}) (and may  be derived from them). 

 The action (\ref{actionA}) gives rise the following equations of motion
\begin{eqnarray}
   \partial_\alpha(E_{[1}^\alpha E_{2]}^\beta)&=&0 ,\label{eqmot1}\\
 \gamma_{\alpha\beta} E_1^\beta - 2\pi\alpha 'F_{\alpha\beta} 
 E_2^\beta & = &0, \label{eqmot2}\\
 \gamma_{\alpha\beta} E_2^\beta - 2\pi\alpha' F_{\alpha\beta} 
 E_1^\beta &= &0, \label{eqmot3}\\
 \partial_\alpha[(E_1^\alpha E_1^\beta - E_2^\alpha E_2^\beta) 
 \partial_\beta X^\mu]& = &0 ,\label{eqmot4}
\end{eqnarray} 
 where for the sake of simplicity we use a flat space-time metric 
 $G_{\mu\nu}=\eta_{\mu\nu}$ (The generalization to a general metric  is
 straightforward).
  In the  gauge $E_1^\alpha = \delta_0^\alpha$  equation (\ref{eqmot1}) reduces to 
\beq
\partial_a E^a_2=0,\,\,\,\,\,\,\,\,\,\,\,\,\,\,\,\,\,
  \partial_0 E_2^a=0. 
\eeq{eqforE2}
 From the action (\ref{actionA}), the canonical 
 momentum $\pi^a$ conjugated to $A_a$ is $- 2\pi\alpha'E_2^a$ and therefore there is
  a constraint $E_2^0=0$. 
 In the present gauge the $2p$ equations (\ref{eqmot2})-(\ref{eqmot3}) reduce to
 $p$ independent equations\footnote{Modulo the linear dependent 
 equations and $F_{0a}$.} (constraints) 
\beq
 \gamma_{0a}-2\pi\alpha' F_{ab}E_2^b=0,\,\,\,\,\,\,\,\,\,\,\,\,\,\,\,
 \gamma_{00} + E_2^a \gamma_{ab} E_2^b = 0.
\eeq{modeqmot}
 These constraints correspond to residual symmetries left after 
 gauge fixing. Also from the action (\ref{actionA}) the canonical 
 momentum conjugated to $A_a$ is $-2\pi\alpha' E_2^a$ (i.e. $\pi^a$) and 
 that the canonical 
 momentum conjugated to $X^\mu$ is $\dot{X}^\mu$ (i.e. $P_\mu$). Thus the 
 constraints (\ref{modeqmot}) coincide with those we have discussed previously
 (up to factor a $2\pi\alpha'$). 
 The equation (\ref{eqmot4}) becomes
\beq
 \partial_0^2X^\mu - E_2^a \partial _a E_2^b \partial_b X^\mu =0.
\eeq{neweqmot4}
 Now we can analyze the solutions of the equations of motion in the given gauge.
  We see that there are no dynamical equations of motion for the ``electric'' 
 $E^a_2$ and magnetic $F_{ab}$ fields. There are only a Gauss' law for $E_2^a$
  and Bianchi identities for $F_{ab}$. As a first example, let us take 
 $E^a_2=(E,0,...,0)$ with $E$  constant and $F_{ab}=0$ and make 
 the following ans\"atz for the solution
 $X^\mu(x_0,x_1, x_2,...,x_p) = Y^\mu(x_0, x_1) f(x_2,...,x_p)$.
 Because of (\ref{modeqmot}) and (\ref{neweqmot4}) we see that $Y^\mu$'s
 satisfy the following equations
\beq
 \dot{Y}^\mu Y'_\mu =0,\,\,\,\,\,\,\,\,\,
 \dot{Y}^\mu \dot{Y}_\mu + E^2 Y'^\mu Y'_\mu=0,\,\,\,\,\,\,\,\,\,
 (\partial_0-E\partial_1)(\partial_0+E\partial_1) Y^\mu =0,
\eeq{Yeqmot}
 where $\dot{Y}^\mu\equiv \partial_0 Y^\mu$ and  $Y'^\mu \equiv \partial_1 Y^\mu$. 
 The function $f$ should satisfy
 the following $(p-1)$ equations
\beq
 (\dot{Y}^\mu Y_\mu) (f \partial_a f)=0,\,\,\,\,\,\,\,\,a=2,...,p. 
\eeq{feqwithoutB}
 As a result of (\ref{Yeqmot}) $Y^\mu$ can be interpreted as a string solution
  in the conformal gauge,  with tension
 $|E|$. The equations (\ref{feqwithoutB}) are solved by 
 requiring $f=const$. Thus the string solutions are completely delocalized in the 
 world-volume coordinates $(x_2,...,x_p)$. In other words, the solution corresponds to
 a set of strings distributed uniformly in the ``transverse directions''
  $(x_2,...,x_p)$. 

 If we now take the same ``electric'' field $E^a_2$ 
 as before but a magnetic field $F_{ab}$
 different from zero, then the equation (\ref{Yeqmot}) stays the same while equation
 (\ref{feqwithoutB}) gets modified to
\beq
(\dot{Y}^\mu Y_\mu) (f \partial_a f)= 2\pi\alpha'E F_{a1},\,\,\,\,\,\,\,\,a=2,...,p. 
\eeq{feqwithB}
 This equation is equivalent to
\beq
 \partial_a (f^2) = \frac{4\alpha'E}{rv} \int\limits_{0}^{\pi} dx_1 F_{a1},
\eeq{feqwithBmod}
 where we  assume that $x_1 \in [0,\pi]$ and $r^2\equiv\frac{1}{\pi}
 \int\limits_{0}^{\pi} dx_1\, Y^\mu Y_\mu$,
  $v\equiv\dot{r}$. Writing 
  $F_{a1}=\partial_{[a}A_{1]}$ one may solve 
 this equation explicitly.   
 The solution $f$ of this equation will define how the string world-sheets
 are distributed  in the directions $(x_2,...,x_p)$. 

 So far we have discussed solutions which have an interpretation as a collection 
 of strings filling the world-volume of the brane. 
 One can also construct 
  solutions which correspond to a single string world-sheet which is completely
 localized in the transverse directions (e.g., $f= \delta(x_2)...\delta(x_p)$ in
  the previous 
 example). However these solutions are singular and  require highly singular 
 electric or magnetic configurations which may be problematic from a computational 
 point of view.
 These solutions are limiting cases of regular
 solutions of the theory (in contemporary parlance, they correspond to the boundary
 points of the moduli space of solutions.).

 The  analysis can be extended along similar lines for other configurations
 of $E^a_1$ and $F_{ab}$. As result  one  sees that
 any classical solution of string theory (Polyakov's action) in 
 conformal gauge can be naturally embedded into the present theory in the given gauge. 
 The details  of the embedding are governed by $E^a_2$ and $F_{ab}$. 
 In a different setup (static gauge) similar results were obtained by Sen 
 \cite{Sen:2000kd}. 
 However not all solutions of the classical equations of motions
 are string-like excitations. For instance, assuming that $X^\mu$ is independent
 of $x_1$ (for the case $E_2^a=(E,0,...,0)$ and $F_{ab}=0$) we find 
 a gauge fixed tensionless $(p-1)$-brane solution, other ans\"atzes
 give point particle solutions etc.      
 
 In quantizing the system one may adopt the same approach as 
 in the previous classical consideration, i.e. treat the $U(1)$ degrees 
 of freedom as background fields.
 Thus one may choose specific configurations of $E^a_2$ and $F_{ab}$ 
 satisfying  Gauss' law and the Bianchi identities and then quantize the 
 system considering only the $X^\mu$ as quantum excitations. Within this
  semiclassical treatment the positive modes of the  $(p+1)$ constraints 
 (\ref{modeqmot}) should be 
 imposed on the physical states. These
 constraints generate a closed algebra  similar to (\ref{LLplus})-(\ref{HHH})
 but not the same (special care is needed for (\ref{HL})). The algebra is
 different because  $E_2^a$ and $A_a$ are not regarded
 as a quantum canonical pair of operators, they are just fixed classical 
 background fields. It is not to  be expected that this way of quantizing is 
 reliable in all regimes of the theory. However it might give a first insight
 into the theory and technically it is straightforward to carry out since 
 the algebra of constraints is a Lie algebra. The case of $E_2^a=0$, $F_{ab}=0$
 corresponds to a tensionless $p$-brane and using the BRST approach this has been
 quantized  previously \cite{Saltsidis:1997nx} (no restrictions such as critical 
 dimensions
 were found). The next natural generalization is
  the case of a non-zero electric field $E_a^2$ and zero magnetic field $F_{ab}$.
 This case should be non-trivial since the Virasoro algebra is contained in
  the full algebra of constraints.  
   
\section{Background fields}

In this section we would like to consider two related questions:
 the vacuum fluctuations and the effects of antisymmetric background fields. 

 Let us begin with a comment on the role of the $B$-field in the tachyonic 
 vacuum. The effect
 of the $B$-field comes from the replacement of $F_{\alpha\beta}$ by  
 ${\cal F}_{\alpha\beta} = \partial_{[\alpha} A_{\beta]} 
 + \partial_a X^\mu  B_{\mu\nu} \partial_\beta X^\nu$ in (\ref{b1}). 
 In the Hamiltonian formalism this results in a redefinition of the momenta 
 $P_\mu$
\beq
 P_\mu \rightarrow P_\mu - B_{\mu\nu}\pi^a \partial_a X^\nu. 
\eeq{Bp} 
With this replacement all expressions in Section 2  are still correct.
  The new string-like constraints $Q^{\pm}$  correspond
 to  string constraints in a non-trivial $B$-field background. Thus
 the effect of the $B$-field is rather trivial.

 Now let us turn to the RR fields.
 When the tachyon is frozen at the vacuum all RR background field decouple 
 completely from the theory. In general the tachyon $T$ is a world-volume degree of
 freedom and has a corresponding kinetic term.
 Proposals for the effective action including a tachyon kinetic term 
 have been put forward in \cite{Garousi:2000tr, Bergshoeff:2000dq, Kluson:2000iy}.
 However in the present discussion (and as is often the practice) we ignore 
 the dynamics of the tachyon field itself 
 and consider $T$ as a background field. Hence the action has the form
$$ S= T_p  \int d^{p+1} x \,\, V(T)
 \sqrt{-det(\gamma_{\alpha\beta} + 2\pi\alpha' F_{\alpha\beta})} + $$
\beq 
+ \mu_p  \int C \wedge dT \wedge e^{2\pi\alpha'F} + O((\partial T)^2)
\eeq{actCS}
 where $C$ is the sum of RR forms. Assuming the expansion\footnote{The 
 expansion (\ref{expanV}) is not always valid, for instance not 
 for $V(T)=(T+1)e^{-T}$ around $T=\infty$. However in this case the argument 
 still goes through,
 since as a result of (\ref{exaction}) the correction to 
 the expansion (\ref{actCT}) is
  $O((\partial T)^2, (V(T))^2)$ (i.e., exponentially small around $T=\infty$).}
 of the  tachyon potential around the vacuum 
\beq
 V(T)= V(T_0) + \frac{1}{2} V''(T_0) (T-T_0)^2 + O(T^3)
\eeq{expanV}
 we  rewrite the action as 
$$ S=  \int d^{p}x\, [(E_1^\alpha E_1^\beta - E_2^\alpha E_2^\beta)
 \gamma_{\alpha\beta} + 2\pi\alpha' E_{[2}^\alpha E_{1]}^\beta F_{\alpha\beta}] + $$
\beq
+  \mu_p  \int C \wedge dT \wedge e^{2\pi\alpha'F} + O((\partial T)^2, T^2).
\eeq{actCT}
 We may regard $T$ as a static configuration interpolating from
 one vacuum to another, like a kink or a vertex. The point is that
 the energy density of these configurations is localized in some of the spatial 
 world-volume coordinates. In an extreme situation 
 $dT$ is a $\delta$-function along those directions. Thus in the  
 approximation used the effect of the Wess-Zumino term is to insert $\delta$ sources on
 the right hand side of (\ref{eqmot1}) and (\ref{eqmot4}). In the gauge 
 $E_2^\alpha = \delta^\alpha_0$ the equation (\ref{modeqmot}) stays the same 
 (thus the Virasoro subalgebra is still present). Apart from those sources
 all the discussion in the previous section goes through. The presence of 
 $\delta$-source on the right hand side of (\ref{eqforE2}) will allow string-like 
 solutions to end where the $\delta$-source sits
 (since it is a source for the flux $\partial_a E^a_2=j_a$). 
 There are thus open strings which can end on ``planes''  
localized in some of the spatial world-volume coordinates, i.e. 
 on lower dimensional D-branes. 

 It is far from obvious that one can drop terms of order 
 $O((\partial T)^2, T^2)$ when discussing  fluctuations around the vacuum.
 Most likely that one cannot freely do so. Nevertheless  the above qualitative
 picture is quite reasonable and agrees with expectations.
 To study the fluctuations around the vacuum  more carefully the
 tachyon field should be treated as dynamical. 

\section{Discussion}

 In the present note we have analyzed the classical effective theory of D-branes at 
 the tachyonic vacuum. 
 We have established the following two facts: the string gauge
 algebra (the Virasoro algebra) is a subalgebra of the D-brane theory at 
the tachyonic vacuum and
  the classical string solutions is a subset of the D-brane  solutions
 in this regime. Thus 
 the Nambu-Goto strings can be embedded into D-branes at the tachyonic 
 vacuum. However, it is not clear to what extent the string picture
 suffices to describe the D-branes in this regime.
 In the similar situation when $\alpha' \rightarrow \infty$ in 
 the fundamental string the quantum theory describes (collection of) massless particles
 \cite{ulf}. By analogy one might expect the quantum theory here to describe 
 (a collection of) strings. In fact, the analogy goes even further than indicated,
 since constraint algebra of tensionless string also contains an ideal 
 ($P_\mu G^{\mu\nu}P_\nu=0$).  
 
 We have emphasized that there are problems with quantizing (\ref{actionA}), but
 the study of the gauge algebra leads to a suggestion
 for quantizing
 the system treating the $U(1)$ degrees of freedom as background degrees of freedom.

\bigskip

\bigskip

{\bf Acknowledgement}: MZ is grateful to Gary Gibbons for valuable discussion.
 We thank Inegemar Bengtsson for the comments on the manuscript. 
 The work of UL was supported in part by NFR grant 650-1998368 and by
 EU contract HPRN-CT-2000-0122.

\appendix
\section{Appendix}

The Lie derivative of arbitrary tensor density of weight $n$
 in the direction of the field $N^a$, is defined as
$$ {\cal L}_{\vec{N}}
T_{a_1 a_2 ... a_k}^{\,\,\,\,\,\,\,\,\,\,\,\,\,\,\,\,\,\,\,
b_1 b_2 ... b_l} =
 N^c \partial_c T_{a_1 a_2 ... a_k}^{\,\,\,\,\,\,\,\,\,\,\,\,\,\,\,\,\,\,\,
b_1 b_2 ... b_l} + \partial_{a_1} N^{c} 
T_{c a_2... a_k}^{\,\,\,\,\,\,\,\,\,\,\,\,\,\,\,\,\,\,\,
b_1 b_2 ... b_l} + ... - $$
\beq
-\partial_c N^{j_1} 
T_{a_1 a_2 ... a_k}^{\,\,\,\,\,\,\,\,\,\,\,\,\,\,\,\,\,\,\,
c b_2... b_l} -... + n \partial_c N^c 
T_{a_1 a_2 ... a_k}^{\,\,\,\,\,\,\,\,\,\,\,\,\,\,\,\,\,\,\,
b_1 b_2 ... b_l}
\eeq{ap7}
We are using the following basic Poisson brackets
\beq
\{ A_{a} (x), \pi^b(y) \} = \delta_a^b \delta^{(p)}(x-y),\,\,\,\,\,\,\,\,\,\,\,\,\,\,
\{X^\mu(x), P_\nu (y) \} = \delta^\mu_\nu \delta^{(p)}(x-y).
\eeq{ap8}
 There is the following action of momentum constraint on the fields
\beq
\{ X^\mu, {\cal H}_a[N^a] \} = {\cal L}_{\vec{N}} X^\mu,\,\,\,\,\,\,\,\,\,\,\,\,\,
\{ P_\mu, {\cal H}_a[N^a] \} = {\cal L}_{\vec{N}} P_\mu,
\eeq{momX}
\beq
\{ F_{ab}, {\cal H}_c[N^c] \} = {\cal L}_{\vec{N}} F_{ab}  ,\,\,\,\,\,\,\,\,\,\,\,\,
\{\pi^a, {\cal H}_c[N^c] \} = {\cal L}_{\vec{N}} \pi^a - N^a \partial_c \pi^c
\eeq{monA}


\begin{thebibliography}{6666}

\newcommand{\np}{{\em Nucl.\ Phys.\ }}
\newcommand{\pr}{{\em Phys.\ Rev.\ }}
\newcommand{\cmp}{{\em Commun.\ Math.\ Phys.\ }}
\newcommand{\pl}{{\em Phys.\ Lett.\ }}

\bibitem{Halpern}
K.~Bardakci, 
" Dual Models and Spontaneous Symmetry Breaking", 
Nucl.\ Phys.\ {\bf B68} (1974) 331;\\
K.~Bardakci and M.B.~Halpern, 
"Explicit Spontaneous Breakdown in a Dual
Model", 
Phys.\ Rev.\ {\bf D10} (1974) 4230;\\
K.~Bardakci and M.B.~Halpern, 
 "Explicit Spontaneous Breakdown in a Dual
Model II: N Point Functions", 
Nucl.\ Phys.\ {\bf B96} (1975) 285;\\
K.~Bardakci, "Spontaneous Symmetry Breakdown in the Standard Dual String
Model", 
Nucl.\ Phys.\ {\bf B133} (1978) 297.

\bibitem{Sen:1999mh}
A.~Sen,
``Descent relations among bosonic D-branes,''
Int.\ J.\ Mod.\ Phys.\  {\bf A14} (1999) 4061
[hep-th/9902105].

\bibitem{Sen:1999md}
A.~Sen,
``Supersymmetric world-volume action for non-BPS D-branes,''
JHEP {\bf 9910} (1999) 008
[hep-th/9909062].

\bibitem{yi}
P.~Yi,
``Membranes from five-branes and fundamental strings from Dp branes,''
Nucl.\ Phys.\ {\bf B550} (1999) 214
[hep-th/9901159].

\bibitem{Bergman:2000xf}
O.~Bergman, K.~Hori and P.~Yi,
``Confinement on the brane,''
Nucl.\ Phys.\  {\bf B580} (2000) 289
[hep-th/0002223].

\bibitem{Harvey:2000jt}
J.~A.~Harvey, P.~Kraus, F.~Larsen and E.~J.~Martinec,
``D-branes and strings as non-commutative solitons,''
JHEP{\bf 0007} (2000) 042
[hep-th/0005031].

\bibitem{Gibbons:2000hf}
G.~Gibbons, K.~Hori and P.~Yi,
``String fluid from unstable D-branes,''
hep-th/0009061.

\bibitem{Sen:2000kd}
A.~Sen,
 ``Fundamental Strings in Open String Theory at the Tachyonic Vacuum,''
hep-th/0010240.

\bibitem{Lindstrom:1997uj}
U.~Lindstr\"om and R.~von Unge,
``A picture of D-branes at strong coupling,''
Phys.\ Lett.\  {\bf B403} (1997) 233
[hep-th/9704051].

\bibitem{Gustafsson:1998ej}
H.~Gustafsson and U.~Lindstr\"om,
``A picture of D-branes at strong coupling. II: Spinning partons,''
Phys.\ Lett.\  {\bf B440} (1998) 43
[hep-th/9807064].

\bibitem{Lindstrom:1999tk}
U.~Lindstr\"om, M.~Zabzine and A.~Zheltukhin,
``Limits of the D-brane action,''
JHEP {\bf 9912} (1999) 016
[hep-th/9910159].

\bibitem{Henneaux:1983um}
M.~Henneaux,
``Transition Amplitude In The Quantum Theory Of The Relativistic Membrane,''
Phys.\ Lett.\  {\bf B120} (1983) 179.

\bibitem{Saltsidis:1997nx}
P.~Saltsidis,
``Tensionless p-branes with manifest conformal invariance,''
Phys.\ Lett.\ {\bf B401} (1997) 21
[hep-th/9702081].

\bibitem{Garousi:2000tr}
M.~R.~Garousi,
``Tachyon couplings on non-BPS D-branes and Dirac-Born-Infeld action,''
Nucl.\ Phys.\  {\bf B584} (2000) 284
[hep-th/0003122].

\bibitem{Bergshoeff:2000dq}
E.~A.~Bergshoeff, M.~de Roo, T.~C.~de Wit, E.~Eyras and S.~Panda,
``T-duality and actions for non-BPS D-branes,''
JHEP {\bf 0005} (2000) 009
[hep-th/0003221].

\bibitem{Kluson:2000iy}
J.~Kluson,
``Proposal for non-BPS D-brane action,''
Phys.\ Rev.\  {\bf D62} (2000) 126003
[hep-th/0004106].

\bibitem{ulf}
A.~Karlhede and U.~Lindstr\"om,
``The Classical Bosonic String In The Zero Tension Limit,''
Class.\ Quant.\ Grav.\ {\bf 3} (1986) L73;\\
J.~Isberg, U.~Lindstr\"om and B.~Sundborg,
``Space-time symmetries of quantized tensionless strings,''
Phys.\ Lett.\ {\bf B293} (1992) 321
[hep-th/9207005].

\end{thebibliography}
\end{document}